\documentclass[twocolumn,prb,superscriptaddress,preprintnumbers,amsmath,amssymb,showpacs]{revtex4}

\usepackage{graphicx}
\usepackage{dcolumn}
\usepackage{bm}

\begin{document}

\def\mathbi#1{\textbf{\em #1}}

\title{Electronic structure and self energies of randomly substituted solids using density functional theory and model calculations}

\author{M. W. Haverkort}
  \affiliation{Max Planck Institute for Solid State Research, Heisenbergstra{\ss}e 1, D-70569 Stuttgart Germany}
\author{I. S. Elfimov}
  \affiliation{Advanced Materials and Process Engineering Laboratory, University of British Columbia, Vancouver, British Columbia V6T 1Z4, Canada}
\author{G. A. Sawatzky}
  \affiliation{Department of Physics and Astronomy, University of British Columbia, Vancouver, British Columbia V6T 1Z1, Canada}

\date{\today}

\begin{abstract}

We describe procedures to obtain the electronic structure of disordered systems using either tight binding like models or quite directly from \textit{ab inito} density functional band structure calculations. The band structure is calculated using super cells much larger than those containing a single minority component atom. We average over a large number of different super cell calculations containing different randomly positioned minority component atoms in the super cell as well as a varying total number of minority component atoms, weighted by the statistical probability. We develop an efficient and simple algorithm for unfolding of these bands, based on Bloch's theorem. The unfolded band-structure obtained in this way exhibits momentum and energy broadened structures replacing the gaps observed in often used single super cell calculations. Using the super cell averaged band-structure one can introduce a self-energy, resulting from the scattering of randomly positioned alloy components. The self-energy is causal, and shows strong energy and some momentum dependence. The self-energy shows rather non-trivial behavior and is in general non-zero at the Fermi-energy, resulting in an ill or undefined Fermi surface. The real-part of the self-energy at the Fermi-energy relates to an apparent violation of Luttinger's theorem. There is no simple relation between the apparent Fermi-surface volume and the electron count. Examples introducing these effects both for model and real binary alloy systems are presented.

\end{abstract}

\pacs{71.15.-m, 71.23.-k, 79.60.Ht, 71.20.Gj, 71.20.Be, 71.20.Eh}
% 71.15.-m 	Methods of electronic structure calculations
% 71.23.-k 	Electronic structure of disordered solids
% 79.60.Ht 	Disordered structures        - Photoemission and photoelectron spectra
% 71.20.Gj 	Other metals and alloys      - Electron density of states and band structure of crystalline solids
% 71.20.Be 	Transition metals and alloys - Electron density of states and band structure of crystalline solids
% 71.20.Eh 	Rare earth metals and alloys - Electron density of states and band structure of crystalline solids

\maketitle

The material specific properties of many systems depend crucially upon chemical substitution or doping, i.e. the intentionally introduction of impurities in an otherwise pure crystal structure. Substitutions can be chosen to introduce or change, the number of charge carriers, local magnetic moments, specific optical properties, or vortex pinning centers, just to mention a few commonly known applications. These substituted atoms also introduce a destruction of translational symmetry and therefore scattering into the band structure as an important deviation from the pure system. The effects of disorder resulting from random substitutions alter material properties in specific ways and the vast field of materials applications is strongly dependent on random substitutions.\cite{James52, Ling88, Bouhafs95, Orner97, Persson01, Maier02, Zorman05, Kakehashi10, Alam10a, Alam10b, Wadati10, Ku10, Berlijn11, Konbu11, Nakamura11, Marmodoro11} 

Although we have powerful \textit{ab initio} band structure methods to treat pure transitionally symmetric systems at least for systems in which electron correlation effects are less important these methods are generally not applicable directly to randomly substituted systems or disordered alloys. There are several useful approximation schemes available to treat randomly substituted systems or alloys, which can treat the disorder to a high level of accuracy.\cite{Faulkner82, Klauder61, Jarrell01,Janis01, Laad05, Batt06, Potthoff07, Kodderitzsch07, Rowlands09, Marodoro11} It is desirable to create a method able to describe disordered systems with the same level of understanding and approximations as one has developed for periodic systems. Density Functional Theory (DFT)\cite{Hohenberg1964}, within the Local Density Approximation (LDA) is an \textit{ab-initio} method proved to be useful for a large class of materials. As long as the materials under consideration are not too correlated, the one electron wave functions in a Kohn and Sham\cite{Kohn1965} implementation can account for the material specific band-structure. First principle methods used for the description of crystals use the periodicity of the system in order to make calculations tractable. Two distinct problems arise when DFT is used to study the electronic structure of real materials: 1) how to account for disorder breaking the translational symmetry and 2) how to present the results in a meaningful way. 

Often, \textit{ab-initio} calculations including disorder are done assuming periodic boundary conditions within a so-call super cell approximation, i.e. using a cell which is a multiple of the primitive unit cell. It is also typical to choose the smallest cell size possible, containing one impurity atom, in order to reduce computational costs. Of course, the smallest cell with only one impurity or dopant, combined with periodic boundary conditions, can not account for a random impurity distribution in real materials. Such calculations basically result in the band structure of a fictitious new perfectly ordered material. Although in some cases such as the extreme dilute limit where the impurity impurity interference effects can be neglected this may be a reasonable approximation to the real problem it is important to develop theoretical methods based on the same approaches and approximations used for pure systems to check for the importance of randomness. For a detailed discussion of development in \textit{ab-initio} theory of alloys we refer to a recent review paper by Ruban and Abrikosov.\cite{Ruban2008} We too use the super cell approximation but allow for randomness in the impurity distribution across the super cell. We then propose a simple procedure to present the unfolded calculated band structures in a way as seen by angle resolved photo emission spectroscopy (ARPES) and demonstrate the effect of impurity scattering on the band structure and the self energy.

We first present the method used to unfold the band-structure in the first Brillouin zone of the super cell to the first Brillouin zone of the original cell. Which is basically based on Bloch's theorem and symmetrization of the wave-functions. The method can be readily applied to all first-principle codes able to calculate partial densities of states and local symmetry or orbital projected band structures. We then demonstrate the basic concepts of the random impurity calculations using a simple tight binding like model and on site impurity potentials. In the last section before the conclusion we describe the use of these concepts in the DFT calculation of a real alloy system composed of Li substituted with either H or Mg.

\section{Unfolding and randomness of impurities}

 \begin{figure}
    \includegraphics[width=0.5\textwidth]{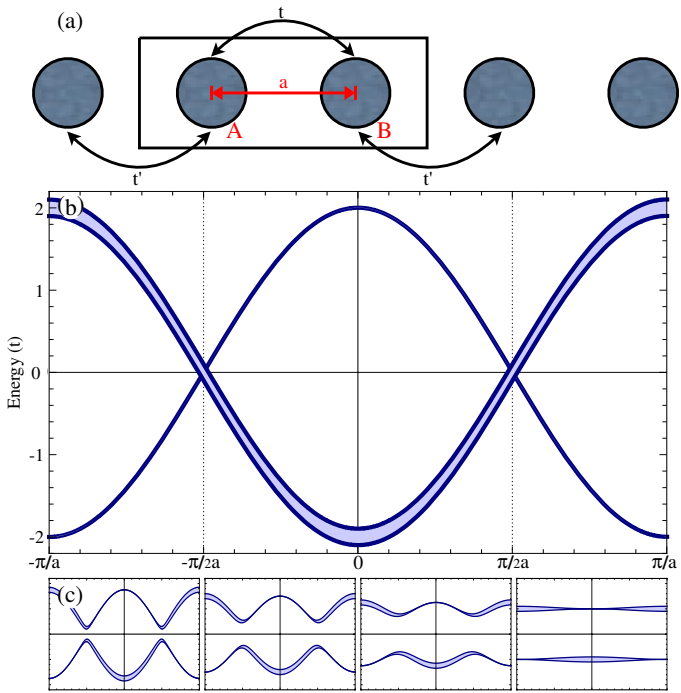}
    \caption{(color online) (a) Tight binding model for an infinite chain of dimers with intra-dimer hopping $t$ and inter-dimer hopping parameter $t'$. (b) Band-structure of the model described in panel (a) with $t'=t$ and the fatness of the bands calculated according to Eq. (\ref{EQunfolding}) scaled by a factor of 0.2 for clarity. The doted lines are the dimer BZ. (c) same as panel (b) but now for (from left to right) $t'=0.75t$, $t'=0.50t$, $t'=0.25t$, $t'=0.00t$.}
    \label{FigDimerExample}
 \end{figure}

In order to recover the original band structure from a super cell calculation one needs to unfold the bands back to the first Brillouin zone of the original cell. Several methods are available, which either look at the momentum dependent Hamiltonian,\cite{Pavarini05} or at the eigenfunctions.\cite{Ku10} In case of some additional distortion, i.e. the super cell is not a perfect repetition of the original primitive-cell unfolding can become a rather non-trivial task. A possible method to define unfolded bands is to assign a certain weight to each eigenstate so that the eigenstates with a weight of 1 represent the original band structure while the eigenstates with a weight of zero identify the folded bands. In this section we first demonstrate how this can be accomplished with a rather simple procedure, after the example we present a more thorough derivation. 

Consider a one-dimensional tight binding model with one s orbital per site and a nearest neighbor hopping integral $t$. The solution to this problem is the wave function that is symmetric at the zone center and antisymmetric at the zone boundary:
\begin{eqnarray}
\Psi(k)=\frac{1}{\sqrt{N}} \sum_{i=1}^{N} e^{\imath kr_i} \phi_i,\\
\nonumber E(k)= 2 t \cos(k a),
\end{eqnarray}
where $t<0$.

As a super cell we take a doubling of the unit cell. This could model an infinite chain of dimers with intra-dimer hopping $t$ and inter-dimer hopping parameter $t'$. Assuming equal on-site energies and constant inter-atomic distances of size $a$, the matrix representing the Hamiltonian is given by:
\begin{equation}
H=
\begin{pmatrix}
     0 & t  + t' e^{-2\imath k a}   \\
     t  + t' e^{2\imath k a} & 0  
\end{pmatrix}.
\end{equation}
The Hamiltonian can be understood by looking at Fig. (\ref{FigDimerExample}). There are two sites within the unit-cell, labeled $A$ and $B$. (The unit cell, indicated by the black box has length $2a$.) Starting from the left site in the unit-cell at the atom labeled $A$ one can hop to the right with the strength $t$ or to the left with strength $t'$. Both processes represent a hopping from site $A$ to site $B$, and thus enter as off-diagonal elements between $\phi_A$ and $\phi_B$.  In the Hamiltonian the hopping is multiplied by a momentum and directional dependent phase, whenever the hopping connects two species in a different unit-cell. For a hopping crossing the super cell boundary the additional phase is $e^{-\imath k R}$ for a hopping to the right, or $e^{+\imath k R}$ for a hopping to the left. With $R=2 a$, the size of the super cell. For $t=t'$ the eigenvalues and the eigen-vectors are:
\begin{eqnarray}
\phi(k)=\sqrt{\frac{1}{2}}(\phi_A\pm e^{\imath k a}\phi_B)\\
\nonumber E(k)=\pm2t \cos(ka),
\end{eqnarray}
The main band has the energy of $+2t$ (remember that $t<0$) at the zone center and $-2t$ at the zone boundary. The wave-function of the unfolded band within the super cell consists of the sum of $\phi_A$ and $\phi_B$ at the zone center and the difference at the zone boundary ($k=\pi/a$). The second solution, i.e. the folded replica, shows the opposite behavior in energy and its wave-function at the zone center (boundary) is made by the difference (sum) of $\phi_A$ and $\phi_B$. Both bands can be seen in panel (b) of Fig. (\ref{FigDimerExample}).

In a true multi-component system one often uses a so-called 'fat' band representation in order to separate various orbital contributions into the band structure at given energy and $\mathbf{k}$-point. There the weights (i.e. $\mathbf{k}$ dependent band thickness) are determined by the squares of the eigen-vector components:
\begin{equation}
\label{eqfatbands}
I_{b,\mathbf{k}}^{\tau,\tau'}=\sum_i \left((C_{b,i,\mathbi{k}}^{\tau'})^*(C_{b,i,\mathbi{k}}^{\tau})\right),
\end{equation}
with $b$ the band index, $\tau$ and $\tau'$ some orbital or site index, $C$ the crystal momentum ($\mathbi{k}$) dependent eigen-vector component, and $i$ a sum over equivalent sites within the unit-cell. It is clear that weights calculated using the eigen-vectors and weights as defined above results in a constant weight throughout the whole Brillouin zone for both bands.

A simple change in the definition of the weighting-function allows one to include information about the relative phase in the wave-functions between different sites. If one defines the weight as:
\begin{equation}
\label{EQunfolding}
I_{b,\mathbf{k}}^{\tau,\tau'}=\frac{1}{N_i}\left(\sum_{i=1}^{N_i} C_{b,i,\mathbi{k}}^{\tau'} e^{-\imath \mathbf{k} \cdot \mathbf{r}_i}\right)^*\left(\sum_iC_{b,i,\mathbi{k}}^{\tau} e^{-\imath \mathbf{k} \cdot \mathbf{r}_i}\right),
\end{equation}
one basically changed the sum of the norm-squared of the eigenstate pre-factors into the norm of the Fourier transform of the eigenstate pre-factors. Defining the positions of the atoms such that $r_A=0$ and $r_B=a$ and applying Eq. (\ref{EQunfolding}) to the wave-functions as found in the dimer model immediately shows that the band belonging to the original, unfolded system, with the wave-function $\sqrt{1/2}(\phi_A+e^{\imath k a}\phi_B)$ gets the weight $(1/2)(\sqrt{1/2}+(e^{-\imath k a} e^{\imath k a})\sqrt{1/2})^2=1$. The folded band, with the wave-function $\sqrt{1/2}(\phi_A-e^{\imath k a}\phi_B)$ obtains the weight, $(1/2)(\sqrt{1/2}-(e^{-\imath k a} e^{\imath k a})\sqrt{1/2})^2=0$.

 \begin{figure}
    \includegraphics[width=0.5\textwidth]{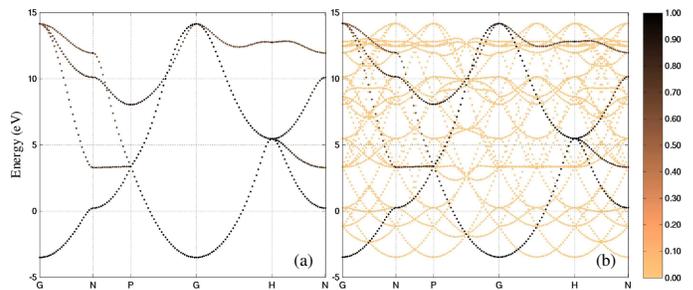}
    \caption{(color online) (a) LDA Li band-structure calculated for the primitive unit cell. (b) Li band-structure calculated for a $2\times2\times2$ super cell. In both panels colored \textit{fat bands} are used to show the $2p$ and $2s$ orbital character within the first Brillouin zone of the primitive unit cell. Unfolding is done with the use of Eq. (\ref{EQunfolding}), traced over the $2p$ and $2s$ orbital character. The zero of energy is at the Fermi energy.}
    \label{FigDFTexample}
 \end{figure}

As a proof of principle, we show in Fig. (\ref{FigDFTexample}) the band structure of Li metal calculated with TB-LMTO density functional code\cite{Andersen75} using a $2\times2\times2$ super cell. The intensity is calculated according to Eq.(\ref{EQunfolding}) in both super cell and primitive cell calculations. Li is a $sp$ metal with a body centered crystal structure and a lattice constant of 3.51 \AA. The primitive-cell contains one Li atom, leading to 4 bands per $\mathbi{k}$-point of different orbital character. At $\Gamma$, for example, the Li $2s$ derived band is at $\approx$-4eV. The three fold degenerate $2p$ bands are at about 14 eV above the Fermi energy. In other regions these bands cross and intermix. The variable $\tau$ labels the orbital character: $\tau\in \{2s, 2p_x, 2p_y, 2p_z\}$. The index $i$ sums over the $16$ equivalent sites within the super cell. The band-index $b$ labels the $4\times16$ different eigenstates found at each $\mathbi{k}$ point within the super cell. To highlight the complete band structure, we use the total intensity, traced over the orbital index $\tau$: $I_{b,\mathbf{k}}^{tot}=\sum_{\tau}I_{b,\mathbf{k}}^{\tau,\tau}$. 

There is a complication with the definition for unfolding presented so-far. In principle one is free to choose the phase of each basis function separately. Multiplying the phase of the basis function in the dimer labeled as $A$ in Fig. (\ref{FigDimerExample}) should not change any physical results. However it does change the unfolded weight as calculated with Eq. (\ref{EQunfolding}). The solution is simple, one should keep track of the phases of the basis functions and introduce a complex-conjugated of this phase in the definition of Eq. (\ref{EQunfolding}). Below we will derive the equation for the weight of an unfolded band-structure more rigorously. The here sketched phase-problem will then come out naturally as additional form-factors in the sum of Eq. (\ref{EQunfolding}).

\subsection{Derivation of general unfolded weight of a band}

The weight of a band can be calculated in similar fashion as one would calculate the momentum (inter-, and intra-Brillouin-zone) depended photo-emission spectra. Neglecting polarization effects, assuming a high photon energy such that the final states can be approximated by plane waves, and neglecting the fact that the action of the electric field of the light on the electrons is given by a momentum operator one can write the spectral weight ($I$) of a one particle eigenstate $\psi$ as: 
\begin{equation}
\label{equnfoldingfourier}
I_{b,\mathbf{k}} \propto \left|\langle e^{\imath \mathbf{k} \cdot \mathbf{r}} | \psi_{b,\mathbi{k}}(\mathbf{r}) \rangle\right|^2,
\end{equation}
whereby $\mathbf{k}$ labels the momentum, $\mathbi{k}$ the corresponding reduced or crystal momentum, and $b$ the band-index. (A momentum dependent mixture of $2s$, $2p_x$, $2p_y$, or $2p_z$ orbital character in the case of Li as shown in Fig. (\ref{FigDFTexample}).)

For a super cell calculation of an undistorted structure it is straight-forward to show that the spectral weight function as defined in Eq. (\ref{equnfoldingfourier}) will be zero for bands that do not belong to the unfolded bands of the primitive unit cell. Take the primitive cell to be defined by the matrix $A$, consisting of three vectors $\mathbf{a}_1$, $\mathbf{a}_2$, and $\mathbf{a}_3$, which span the parallel piped defining the primitive unit cell. Using Bloch's theorem for a periodic system we know:
\begin{equation}
\label{eqBloch}
\psi_{b,\mathbi{k}}(\mathbf{r})=e^{\imath \mathbi{k} \cdot \mathbf{r}} u_{b,\mathbi{k}}(\mathbf{r}),
\end{equation}
with,
\begin{equation}
u_{b,\mathbi{k}}(\mathbf{r}+\mathbf{n}\cdot A)=u_{b,\mathbi{k}}(\mathbf{r}), \quad\quad \forall \mathbf{n}\in\mathbb{Z}^{(3)}.
\end{equation}
Here $\mathbf{n}$ is an integer vector ($\mathbf{n}\cdot A=n_1\mathbf{a}_1+n_2\mathbf{a}_2+n_3\mathbf{a}_3$) and $u_{b,\mathbi{k}}(\mathbf{r})$ is a periodic function with the period of the primitive unit cell, $\psi_{b,\mathbi{k}}(\mathbf{r})$ is periodic up-to a phase:
\begin{equation}
\psi_{b,\mathbi{k}}(\mathbf{r+\mathbf{n}\cdot A})=e^{\imath \mathbi{k} \cdot (\mathbf{n}\cdot A)} \psi_{b,\mathbi{k}}(\mathbf{r}).
\end{equation}

For a calculation done in the primitive unit cell one can use the block diagonal in $\mathbi{k}$ characteristic of the Hamiltonian. Solving $H$ for each $\mathbi{k}$ vector separately yields the band-structure. Within a super-cell calculation one does not obtain the momentum of a wave-function directly. If one takes a super-cell without perturbing the Hamiltonian the eigenstates are non-the-less given by Bloch's theorem as given in Eq. (\ref{eqBloch}). The Fourier transform, of such a function, which defines the spectral weight as defined in Eq. (\ref{equnfoldingfourier}) is given as:
\begin{eqnarray}
\label{unfoldingBlochWave}
I_{b,\mathbf{k}} &\propto&\Big| \int_{\Omega} e^{-\imath \mathbf{k} \cdot \mathbf{r}} \psi_{b,\mathbi{k}}(\mathbf{r}) \, \partial \mathbf{r}\Big|^2\\
\nonumber &=& \Big|\sum_j \int_{\Omega_p} e^{-\imath \mathbf{k} \cdot (\mathbf{r}+\mathbf{n}_j \cdot A)} \psi_{b,\mathbi{k}}(\mathbf{r}+\mathbf{n}_j \cdot A) \, \partial \mathbf{r}\Big|^2\\
\nonumber &=& \Big|\sum_j \int_{\Omega_p} e^{-\imath \mathbf{k} \cdot (\mathbf{r}+\mathbf{n}_j \cdot A)} e^{\imath \mathbi{k} \cdot (\mathbf{r}+\mathbf{n}_j \cdot A)} u_{b,\mathbi{k}}(\mathbf{r}) \, \partial \mathbf{r}\Big|^2\\
\nonumber &=& \Big|\sum_j e^{\imath (\mathbi{k}-\mathbf{k}) \cdot (\mathbf{n}_j \cdot A)}\Big|^2\Big|\int_{\Omega_p} e^{\imath (\mathbi{k}-\mathbf{k}) \cdot \mathbf{r}} u_{b,\mathbi{k}}(\mathbf{r}) \, \partial \mathbf{r}\Big|^2,
\end{eqnarray}
with $\mathbi{k}$ the crystal momentum of the wave-function in the primitive unit cell. The reduced momentum $\mathbi{k}$, is an eigenstate property according to Bloch's theorem, but undetermined from the super-cell calculation. The reduced momentum index within the super-cell has been suppressed as it is up to a super-cell reciprocal lattice vector implicitly labeled by $\mathbi{k}$. $\Omega$ labels an integral over all space and $\Omega_p$ an integral over the original primitive unit cell, and the sum over all possible independent shifts by an integer lattice vectors ($\mathbf{n}\cdot A$) of the primitive unit cell. The first term of the unfolded spectral weight in the primitive unit cell ($I_{b,\mathbf{k}}$) contains a sum over all unit-cells. This term leads to the selection rule:
\begin{equation}
\mathbf{k}-\mathbi{k}=\mathbf{n} \cdot G,
\end{equation}
with $\mathbf{n}$ an integer vector and $G$ a matrix representing the reciprocal lattice: $G^{T}=2\pi A^{-1}$. This term is responsible for the fact that only those bands that belong to the principle unit cell will gain weight, i.e. it unfolds the folded band-structure. The second term of the unfolded spectral weight is a material and Brillouin zone dependent form factor. In photo-emission experiments it is this term (together with the here neglected polarization dependence) that is responsible for the Brillouin zone dependent spectral weight.

The unfolding of bands as described in Eq. (\ref{equnfoldingfourier}) can be simplified further by approximating the Brillouin zone dependent form factor. We will first briefly describe model calculations and then extend it to a formalism that can readily, without much effort, be implemented in most \textit{ab initio} density functional programs.

\subsubsection{Unfolding of bands in tight-binding model calculations}

For model calculations one often expresses the band-structure in terms of some form of tight-binding representation. If we take the variable $\tau$ as the index for the local orbital and or spin character of the tight-binding basis set (in the case of Li $\tau$ is $2s$, $2p_x$, $2p_y$, or $2p_z$) then one can define the spectral weight matrix for a given one-particle eigenstate ($\psi_{b,\mathbi{k}}(\mathbf{r})$) as:
\begin{equation}
\label{equnfoldingmodel}
I_{b,\mathbf{k}}^{\tau,\tau'}=\langle \psi_{b,\mathbi{k}}(\mathbf{r}) | a^{\dag}_{\mathbf{k},\tau'} a^{\phantom{\dag}}_{\mathbf{k},\tau} | \psi_{b,\mathbi{k}}(\mathbf{r}) \rangle,
\end{equation}
whereby the operator $a^{\phantom{\dag}}_{\mathbf{k},\tau}$ annihilates an electron at momentum $\mathbf{k}$ and of orbital (spin) character $\tau$. The momentum dependent annihilation operator can be written as a sum of position dependent operators as:
\begin{equation}
a^{\phantom{\dag}}_{\mathbf{k},\tau} = \frac{1}{\sqrt{N_i}}\sum_i e^{\imath \mathbf{k} \cdot \mathbf{r}_i } a^{\phantom{\dag}}_{\mathbf{r}_i,\tau},
\end{equation}
which describes the relation between the Fourier transform of a one-particle wave-function and the expectation value of the momentum dependent occupation number operator. If the orbital character of a certain band is of no importance one can write the total unfolded spectral weight of a certain band as the trace over the spectral weight matrix:
\begin{equation}
I_{b,\mathbf{k}}=\sum_{\tau} I_{b,\mathbf{k}}^{\tau,\tau}.
\end{equation}

\subsubsection{Unfolding of bands for \textit{ab initio} density functional calculations.}

In the case of an \textit{ab initio} density-functional theory calculation one could naturally first create a tight binding model and then use Eq. (\ref{equnfoldingmodel}) in order to calculate the unfolded spectral weight. Such a scenario has recently been proposed by Wei Ku \textit{et al.} \cite{Ku10} and successfully used on several systems.\cite{Berlijn11,Konbu11} The unfolded band-structures found in this way are expected to be rather similar to the unfolded band-structure one obtains by defining the weight of a band via a Fourier transform (see Eq. (\ref{equnfoldingfourier})). For large super cells with several impurity atoms, defining a tight-binding Hamiltonian might become troublesome. An alternative method that allows one to directly obtain the unfolded band-structure from an \textit{ab initio} calculation therefore is useful.

One can simplify the Brillouin zone dependent form factor of the unfolded spectral weight as it appears in Eq. (\ref{unfoldingBlochWave}) by looking at the one-particle wave-function character within a certain (muffin-tin) sphere defined around each atom. These projections are implemented in most \textit{ab initio} DFT codes, where these projections are used to calculate the partial character of a given eigenstate needed to plot partial density of states, or bands of a certain orbital character. The projected wave-function $\widetilde{\psi}_{b,\mathbi{k}}(\mathbf{r})$ can be written as:
\begin{equation}
\widetilde{\psi}_{b,\mathbi{k}}(\mathbf{r})=\frac{1}{\sqrt{N_i}}\sum_{\tau,i} C_{b,i,\mathbi{k}}^{\tau} \phi_{\tau,i}(\mathbf{r}-\mathbf{r}_i),
\end{equation}
with $b$ a band index of the one-particle eigenstates, $\tau$ an orbital and spin index, the sum over all sites $i$ that contain a muffin tin sphere onto which one projects, $\mathbf{r}_i$ the position of the sphere labeled with the index $i$, $\phi_{\tau,i}$ the functions projected onto (with orbital character $\tau$ and centered at position $\mathbf{r}_i$), and $C_{b,i,\mathbi{k}}^{\tau}$ the projected pre-factor. Similarly as done for model calculations it makes sense to define an orbital ($\tau$) specific unfolded weight, starting from:
\begin{equation}
\widetilde{\psi}^{\tau}_{b,\mathbi{k}}(\mathbf{r})=\frac{1}{\sqrt{N_i}}\sum_{i} C_{b,i,\mathbi{k}}^{\tau} \phi_{\tau,i}(\mathbf{r}-\mathbf{r}_i),
\end{equation}
Unfolding such a projected one-particle wave-function with the use of Eq. (\ref{equnfoldingfourier}) yields:
\begin{widetext}
\begin{eqnarray}
I_{b,\mathbf{k}}^{\tau,\tau'} &=&\Big(\int_{\Omega} e^{-\imath \mathbf{k} \cdot \mathbf{r}} \widetilde{\psi}^{\tau'}_{b,\mathbi{k}}(\mathbf{r}) \, \partial \mathbf{r}\Big)^{*}\Big(\int_{\Omega} e^{-\imath \mathbf{k} \cdot \mathbf{r}} \widetilde{\psi}^{\tau}_{b,\mathbi{k}}(\mathbf{r}) \, \partial \mathbf{r}\Big)\\
\nonumber&=&\frac{1}{N_i}\Big(\sum_i\int_{\Omega} e^{-\imath \mathbf{k} \cdot \mathbf{r}} C_{b,i,\mathbi{k}}^{\tau'} \phi_{\tau'}(\mathbf{r}-\mathbf{r}_i) \, \partial \mathbf{r}\Big)^{*}\Big(\sum_i\int_{\Omega} e^{-\imath \mathbf{k} \cdot \mathbf{r}} C_{b,i,\mathbi{k}}^{\tau} \phi_{\tau}(\mathbf{r}-\mathbf{r}_i) \, \partial \mathbf{r}\Big)\\
\nonumber&=&\frac{1}{N_i}\Big(\sum_i e^{-\imath \mathbf{k} \cdot \mathbf{r}_i} C_{b,i,\mathbi{k}}^{\tau'} \int_{\Omega} e^{-\imath \mathbf{k} \cdot \mathbf{r}} \phi_{\tau',i}(\mathbf{r}) \, \partial \mathbf{r}\Big)^{*} \Big(\sum_i e^{-\imath \mathbf{k} \cdot \mathbf{r}_i} C_{b,i,\mathbi{k}}^{\tau} \int_{\Omega} e^{-\imath \mathbf{k} \cdot \mathbf{r}} \phi_{\tau,i}(\mathbf{r}) \, \partial \mathbf{r}\Big).
\end{eqnarray}
\end{widetext}
The integral:
\begin{equation}
\label{eqFormFactor}
\int_{\Omega} e^{-\imath \mathbf{k} \cdot \mathbf{r}} \phi_{\tau,i}(\mathbf{r}) \, \partial \mathbf{r},
\end{equation}
defines a Brillouin zone dependent form factor, and although this is a physical quantity one can simplify the unfolding of bands substantially by neglecting this term. If the orbital basis set is defined such that the main outer part of $\phi_{\tau,i}(\mathbf{r})$ is real and positive for all $\tau$ then a powerful approximation is to set the integral in Eq. (\ref{eqFormFactor}) equal to 1. In this case the unfolded weight becomes:
\begin{equation}
I_{b,\mathbf{k}}^{\tau,\tau'} = \frac{1}{N_i} \Big(\sum_i e^{-\imath \mathbf{k}\cdot r_i} C_{b,i,\mathbi{k}}^{\tau'}\Big)^*\Big(\sum_i e^{-\imath \mathbf{k}\cdot r_i} C_{b,i,\mathbi{k}}^{\tau}\Big).
\end{equation}
In the case that the super cell is undistorted, i.e. a pure repetition of the system one can understand that this definition of the spectral weight indeed unfolds the band-structure with the argumentation based on Bloch's theorem. This can be done by following the argumentation as given in the beginning of this section from Eq. (\ref{eqBloch}) to Eq. (\ref{unfoldingBlochWave}). When the super cell is not a perfect repetition of primitive basis cells then the weight of the folded bands starts to deviate from zero. The weight in the folded bands is related to the mixing of the folded and original bands in the band-structure and therefore a measure of the distortion.

One should be slightly careful how the form-factor as presented in Eq. (\ref{eqFormFactor}) is neglected. The phase of this integral depends on the phase of the basis function $\phi_{\tau,i}(\mathbf{r})$, which can be chosen freely. In many calculations the projections are done onto atomic like wave-functions. These functions are real, which fixes most of the phase, but still can either be positive or negative. The radial part of atomic like wave-functions has nodes. The important outer part of the radial wave-function is either positive or negative, depending on the number of nodes. The number of nodes for an atomic like wave-function is given as $n-l$ the phase of the basis function therefore is $(-1)^{(n-l)}$. Naturally different implementations might use different phases for the basis functions. (A complex phase of $e^{-\imath \mathbf{k}\cdot r_i}$ can for example be a useful alternative choice.) Using our phase definition, including the factor $(-1)^{(n-l)}$ for the basis functions, the Brillouin zone dependent form factor will be approximated as:
\begin{equation}
\int_{\Omega} e^{-\imath \mathbf{k} \cdot \mathbf{r}} \phi_{\tau,i}(\mathbf{r}) \, \partial \mathbf{r} \to (-1)^{(n_{i}^{\tau}-l_{i}^{\tau})},
\end{equation}
whereby $n_{i}^{\tau}$ is the principle quantum number of the orbital with index $\tau$ and position $\mathbf{r}_i$ and $l_{i}^{\tau}$ is the angular momentum of the orbital with index $\tau$ and position $\mathbf{r}_i$. The unfolded weight then becomes:
\begin{eqnarray}
\label{equnfoldingdft}
I_{b,\mathbf{k}}^{\tau,\tau'}&=&\frac{1}{N_i}\Big(\sum_i e^{-\imath \mathbf{k} \cdot \mathbf{r}_i} (-1)^{(n_{i}^{\tau'}-l_{i}^{\tau'})} C_{b,i,\mathbi{k}}^{\tau'} \Big)^* \\
\nonumber&&\qquad\times\Big(\sum_i e^{-\imath \mathbf{k} \cdot \mathbf{r}_i} (-1)^{(n_{i}^{\tau}-l_{i}^{\tau})} C_{b,i,\mathbi{k}}^{\tau} \Big).
\end{eqnarray}
The partial character $C_{b,i,\mathbi{k}}^{\tau}$ is the projected weight of the eigenstate with band index $b$, orbital and spin character $\tau$, at site $i$ in the super cell for the crystal momentum $\mathbi{k}$. The unfolded weight of this band is calculated by taking a Fourier transform summed over (pseudo) equivalent sites ($i$) within the super-cell. This can either be the same atom at positions in the super cell slightly deviating from a perfect repetition of the primitive unit cell, or different atoms in the super cell in the case of substitution. The norm-squared of this Fourier transform defines the unfolded weight of a certain band and orbital character.

\subsection{configuration average for the description of randomness}

 \begin{figure}
    \includegraphics[width=0.5\textwidth]{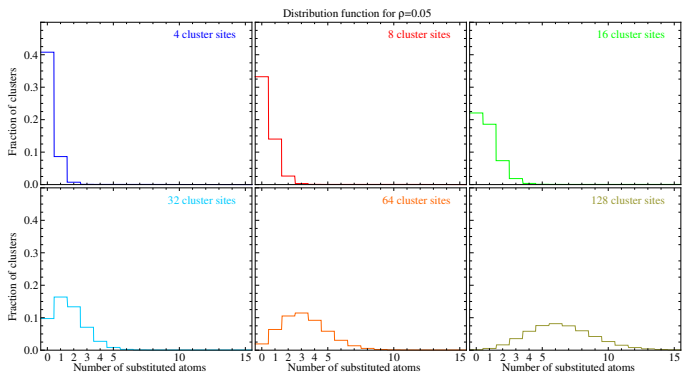}
    \caption{(color online) Statistical fraction of clusters with $N_s$ substituted atoms for a cluster size of 4 (left top) to 128 (right bottom) $N_t$ total atoms and an average substitution of 5\% ($\rho=0.05$).}
    \label{FigDistributionFunction}
 \end{figure}

There is a crucial difference between a calculation for an ordered structure and a system with randomly substituted atoms. In a system with ordered impurities gaps will open in the band-structure, whereas for a random system this may not be the case and gaps can become smeared out structures, as will be shown by several examples later. In order to simulate the situation of an infinite crystal with a random impurity distribution one would like to create very large super cells. In order to do these calculations with the use of tractable cluster sizes it is important to average over different random distributions of impurities with equal impurity density and over different impurity densities within the cluster. Lets define the chance to have a substituted atom at site $i$ to be equal to $\rho$. For a cluster of $N_t$ total sites the probability to have a number of $N_s$ substituted atoms within the cluster is given by the binomial distribution:
\begin{equation}
P(N_{s},N_{t};\rho)=\rho^{N_{s}} (1-\rho)^{(N_t-N_s)} \frac{N_t!}{N_s!(N_t-N_s)!}.
\label{EqDistributionFunction}
\end{equation}
In order to illustrate the impurity distribution further we show in Fig. (\ref{FigDistributionFunction}) a plot of this distribution for a doping of 5\% ($\rho=0.05$) and a cluster size of 4, 8, 16, 32, 64, and 128 $N_t$ total atoms. One should realize that even for a reasonable cluster size there still is a finite chance to have no substituted atoms within the cluster. For each total number of substituted atoms in the cluster there are still many different ways to arrange these impurities. It is important to take this into account, since the scattering of nearest neighbor impurities is in general very different from the scattering of isolated impurities. In general one should determine the band structure in a super cell method with as large a cell as possible in real space and average over a varying number of impurities and a varying distribution of these impurities. Each of these calculations should be weighted with the probability to find this particular kind of configuration, including both the chance to have this specific density within the cluster and this specific order within the cluster. In practice it is not needed to calculate all possible impurity configurations (including variation of density and distribution) within the cluster. A random subset will already give a reasonable approximation to the real solution, provided that the cell is large enough and a large number of random configurations are taken.

We define a configuration to be a specific realization of a super cell with a random number of substituted atoms at random positions. The average density of substituted sites is given by $\rho$. 

\section{Random onsite energy impurity model}

Here we exemplify several aspects of random impurity calculations, using a simple model system. The model we use is well studied which allows us to compare the super cell calculations to theoretical approximations used elsewhere.\cite{Klauder61,Faulkner82,Janis01,Jarrell01,Laad05,Batt06,Rowlands09} We use a single band model with constant nearest neighbor hopping ($t$) and a onsite impurity potential $V$ at a fraction $\rho$ of all atoms randomly placed within the solid. The Hamiltonian is given as:
\begin{equation}
\label{eqModel}
H=\sum_{\langle i,j \rangle} t a^{\dag}_i a^{\phantom{\dag}}_{j} + \sum_{i} V_{i} a^{\dag}_i a^{\phantom{\dag}}_i,
\end{equation}
with the sum over all sites $i$ and nearest neighbor sites $j$. $t$ is the hopping-strength, resulting in a band-width of $W = 2 D t$. $D$ is the dimension of the system. $V_{i}$ is a potential with the value $V$ at a fraction $\rho$ of randomly placed sites and zero at other sites. $\rho$ is the density of substituted atoms. 

As the Hamiltonian only includes one-particle interactions it is natural to Fourier transform it to momentum ($\mathbf{k}$) space:
\begin{equation}
H=\sum_{\mathbf{k}} \epsilon_{\mathbf{k}}^{\phantom{\dag}} a^{\dag}_{\mathbf{k}} a^{\phantom{\dag}}_{\mathbf{k}} + \sum_{\mathbf{k},\mathbf{k'}} V_{\mathbf{k},\mathbf{k'}}^{\phantom{\dag}} a^{\dag}_{\mathbf{k}} a^{\phantom{\dag}}_{\mathbf{k'}},
\end{equation}
with $V_{\mathbf{k},\mathbf{k'}}=\frac{1}{N}\sum_{i} e^{\imath (\mathbf{k}-\mathbf{k'})\cdot\mathbf{r_i}} V_{i}$ and $\mathbf{r_i}$ the position of atom $i$. Within this representation the hopping part becomes diagonal in $\mathbf{k}$. The random potential allows for scattering between different $\mathbf{k}$-vectors. In the limit of large cluster sizes or for a potential averaged over many different cluster configurations one finds that $V_{\mathbf{k},\mathbf{k}}=V\rho$ and that $|V_{\mathbf{k},\mathbf{k'}}|$ for $\mathbf{k}\neq\mathbf{k'}$ scales as $\sqrt{1/N}$. The averaged Hamiltonian ($\overline{H}$), is thus defined as $\overline{H}=\sum_{\mathbf{k}} (\epsilon_{\mathbf{k}}+V\rho) a^{\dag}_{\mathbf{k}} a^{\phantom{\dag}}_{\mathbf{k}}$. One might be tempted to state that for infinitely large clusters or for a cluster of finite size, averaged over a large number of configurations the Hamiltonian becomes diagonal in momentum $\mathbf{k}$ and thus momentum is always a good quantum number. Care has to be taken as there is indeed only an infinitesimal small coupling to different momenta for infinitely large clusters, but there are an infinite number of different momenta possible to which one can scatter  thus leading to a finite perturbation.

The one particle Green function is defined as:
\begin{equation}
G_{\mathbf{k},\mathbf{k'}}(\omega)=\lim_{\Gamma\to0^+}\langle 0 | a^{\phantom{\dag}}_{k'} \frac{1}{\omega-H+\imath \Gamma/2} a^{\dag}_{k} | 0 \rangle.
\end{equation}
The spectral function whose poles define the band-structure, is given as $A(\mathbf{k},\omega)=-\frac{1}{\pi}\mathrm{Im}[G_{\mathbf{k},\mathbf{k}}(\omega)]$. The off diagonal elements of $G_{\mathbf{k},\mathbf{k'}}(\omega)$ in $\mathbf{k}$, i.e. those elements for which $\mathbf{k}\neq\mathbf{k'}$ scale as a function of cluster size, in the same way as the off diagonal elements of the Hamiltonian, namely as $\sqrt{1/N}$. 
For large clusters or for an average over many different random configurations the Green function becomes diagonal in momentum $G_{\mathbf{k}}(\omega)=\overline{G_{\mathbf{k},\mathbf{k'}}(\omega)}$. Calculating the Green function, using the average Hamiltonian (which is diagonal in momentum) is however different from using the full Hamiltonian, even if one averages the Green function afterward. In the following whenever we write $G_{\mathbf{k}}(\omega)$ we will assume that this is the Green function averaged over several configurations ($G_{\mathbf{k}}(\omega)=\overline{G_{\mathbf{k},\mathbf{k'}}(\omega)}$).

\subsection{Moment analyzes and series expansion}

The random onsite energy impurity model has been extensively studied with the use of series expansions.\cite{Klauder61,Faulkner82} The $n$-th moment ($\mu^{(n)}$) of the momentum integrated spectral function (density of states) is given as:
\begin{equation}
\mu^{(n)}=\frac{1}{N} \mathrm{Tr}[H^n],
\end{equation}
with $N$ the total number of sites in the cluster. The Hamiltonian is given as $H^{(0)}+H^{(1)}$ with $H^{(0)}=\sum_{\mathbf{k}} \epsilon_{\mathbf{k}} a^{\dag}_{\mathbf{k}}a^{\phantom{\dag}}_{\mathbf{k}}$ and $H^{(1)}=\sum_{i} V_{i} a^{\dag}_i a^{\phantom{\dag}}_i$. Analytical expressions can immediately be written down, if one neglects contributions to the moment which arise from products of $H^{(0)}$ and $H^{(1)}$. The moments of $H^{(0)}$ are given as:
\begin{equation}
\mu_0^{(n)}=\left(\frac{1}{2\pi}\right)^D\int_{-\pi}^{\pi} \left(\sum_{i=1}^D 2\,t\,\mathrm{sin}[k_i]\right)^n \delta\mathbf{k},
\end{equation}
which are the moments of the original, unperturbed system. The non-zero lowest order moments are $\mu_0^{(0)}=1$, $\mu_0^{(2)}=2 D t^2$, and $\mu_0^{(4)}=6 t^4 D (2D-1)$. The moments of $H^{(1)}$ are:
\begin{equation}
\mu_1^{(n)}=V^n\rho.
\end{equation}
Contributions due to products of $H^{(0)}$ and $H^{(1)}$ are important for the moments with $n\ge3$.

The momentum $\mathbf{k}$ dependent Green function can be expressed in terms of a series in $H^{(1)}$ with the use of the Dyson equations. For a Hamiltonian only containing one electron terms one can write:
\begin{equation}
G_{\mathbf{k},\mathbf{k'}}(\omega)=G_{\mathbf{k}}^{(0)}(\omega)+\sum_{k''}G_{\mathbf{k}}^{(0)}(\omega)H_{\mathbf{k},\mathbf{k''}}^{(1)}G_{\mathbf{k''},\mathbf{k'}}(\omega).
\label{eqDyson}
\end{equation}
For the random configuration averaged Green function, $\overline{G_{\mathbf{k},\mathbf{k'}}(\omega)}=G_{\mathbf{k}}(\omega)$ one finds to first order in the onsite impurity potential $V$:
\begin{equation}
G_{\mathbf{k}}(\omega)=G_{\mathbf{k}}^{(0)}(\omega)+G_{\mathbf{k}}^{(0)}(\omega)V\rho G_{\mathbf{k}}(\omega).
\label{eqGSeries}
\end{equation}
It is often useful to define the self energy for the averaged Green function as:
\begin{equation}
\Sigma(\mathbf{k},\omega)=\frac{1}{G_{\mathbf{k}}^{(0)}(\omega)}-\frac{1}{G_{\mathbf{k}}^{\phantom{(}}(\omega)}.
\label{eqSelfEnergy}
\end{equation}
If one compares the definition of the self energy with Eq. (\ref{eqDyson}) one finds a striking similarity. Multiplying Eq. (\ref{eqSelfEnergy}) with $G_{\mathbf{k}}^{(0)}(\omega)$ and $G_{\mathbf{k}}(\omega)$, one reproduces the form of Eq. (\ref{eqDyson}), but now for the configuration averaged, single momentum dependent Green function, instead of the original Green matrix. Using Eq. (\ref{eqGSeries}) one finds that to first order in $V$ the self energy is given as $\Sigma=V\rho$. To linear order in the perturbing potential one can thus replace the random impurity system by a periodic system, which has an averaged potential on each site. This approximation is known as the virtual crystal approximation. It can correctly describe the averaged band-structure and total density of states to first order in $V$. It can not describe changes to the one-particle wave-functions important for the band-character, the spatially varying charge density,\cite{Wadati10} the broadening of bands, or the possible appearance of impurity bound states, which arise in quadratic and higher order of the perturbing potential.

\subsection{Cluster size and configuration averaging}

 \begin{figure}
    \includegraphics[width=0.5\textwidth]{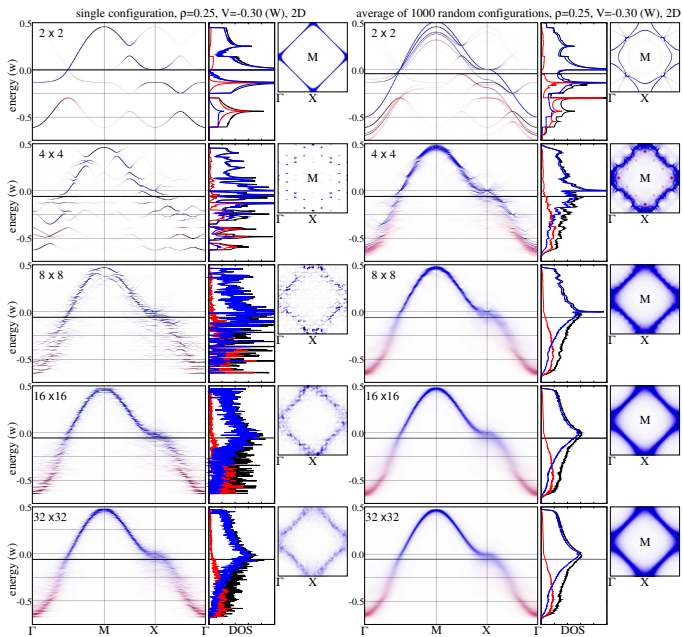}
    \caption{(color online) Unfolded band-structure, density of states and Fermi-surface of a $2D$ tight binding model as presented in Eq. (\ref{eqModel}), for various cluster sizes. 
The left column shows calculations for one random impurity distribution with $\rho=0.25$.
The right column shows calculations for an average of 1000 different random configurations, which only on average have an impurity density of 25\%, and vary both in impurity density as well as impurity distribution within a cluster. In blue (red) the character of the host (substituted) atoms is shown. The potential of the substituted atoms is at $V=-0.3W$, with $W$ the band-width (the potential of the host atoms is at 0). The Fermi-surface plots assume a filling of on average 1 electron per atom.}
    \label{FigClusterSize}
 \end{figure}

In order to fully describe the effects of random impurities it is necessary to go beyond a series expansion in the perturbing potential. For such cases doing exact diagonalization of large clusters is a powerful method, especially with the current computational resources. Within this section we compare several calculations done for different cluster sizes. We furthermore show that the cluster sizes can be considerably reduced if one averages over several random configurations, i.e. both varying the impurity density as well as the impurity distribution within the cluster.

In the left most column of Fig. (\ref{FigClusterSize}) we show calculations for the band-structure of a single random configuration, with fixed concentration. The different rows show the calculation for a different total cluster size. The calculations are done for a 25\% ($\rho=0.25$) substitution and $V=-0.30W$. The smallest cluster possible which allows for a substitution of 25\%, namely with a total of $N_t=4$ sites and one impurity atom, shows the opening of large gaps in the band-structure. This is to be expected as for a $2\times2$ cluster with $1/4$ substitution one has a perfectly ordered super-structure, which is very different from a random system. The larger the cluster becomes the smoother the band-structure becomes. The gaps become smaller and smaller and more and more distributed to random places in momentum space. One however needs very large clusters to find convergence if one does not average over different random configurations. Even for a single calculation of a $32\times32$ super cell there is still some gaping visible in the band-structure (left column bottom row of Fig. (\ref{FigClusterSize})). This becomes much better if one averages over different random configurations (i.e. averaging over both a varying number of substituted atoms within the cluster and averaging over the possible different ways to substituted these atoms). The forth column shows the same band-structure calculations, but now averaged over 1000 random configurations. As one can see these look much more smooth. The gaps disappear when an average over different configurations is taken. Even for a $4\times4$ super cell calculation the averaged band-structure looks quite reasonable, whereas the calculation of a single configuration is far from converged. What happens is that each different random configuration has gaps at different energies and $\mathbf{k}$ vectors, leading to an averaged broadening of the bands and not to a gaping of the bands. A similar effect can be seen for the density of states. The calculations for a single configuration are rather noisy; the average over 1000 configurations however shows a nice continues behavior. 

Averaging is crucial for the calculation of the Fermi-surface. The Fermi-surface is calculated by assuming a filling of 1 electron per atom in the unit cell after averaging all calculations done for different configurations. A single configuration shows only some points on the Fermi-surface, whereas an average over several configurations shows a very nice, though broadened Fermi-surface. Note that in the Fermi-surface plots for small super cell calculations one can see the periodicity of the super cell very clearly as features with the periodicity of the super cell Brillouin zone. The use of small super cell calculations can lead to spurious shadow bands.

\subsection{Comparison to other approximations}

 \begin{figure}
    \includegraphics[width=0.5\textwidth]{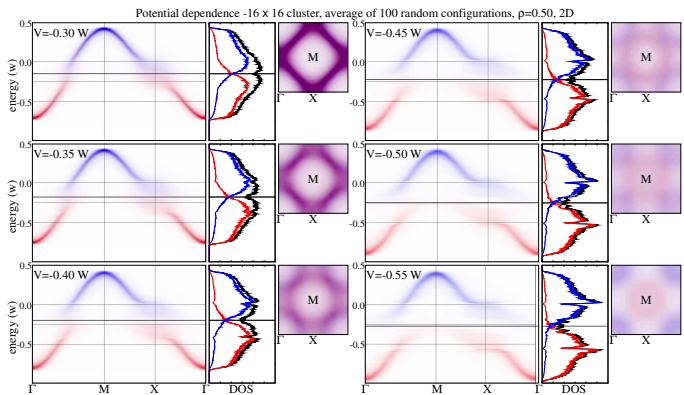}
    \caption{(color online) Unfolded band-structure, density of states and Fermi-surface of a $2D$ model calculation as defined in Eq. (\ref{eqModel}), for a $16\times16$ cluster, with $\rho=0.50$ and $V=-0.30 W$ to $V=-0.55W$. The Fermi-surface plots assume a filling of on average 1 electron per atom.}
    \label{FigPotentialDependence2D}
 \end{figure}

Next we compare the direct method of calculating the band-structure, Fermi-surface and density of states as presented here with other methods like the coherent potential approximation and its non-local version.\cite{Faulkner82,Jarrell01} in Fig. (\ref{FigPotentialDependence2D}) we present calculations as a function of the impurity potential for a $2D$ 50\% ($\rho=0.5$) substituted system. The model and parameter range used are the same as presented in Fig. (5) of the paper by Batt and Rowlands \cite{Batt06}, which allows for a direct comparison between the different approximations. For a potential of $V=-0.3W$ one finds a clear Fermi-surface at a filling of one electron per site. The Fermi-surface is rather broad. These findings are basically equivalent to the results found with the coherent potential approximation and for the non-local coherent potential approximation calculated using a $2\times2$ cluster. Increasing the impurity potential further will eventually lead to two separate bands separated by a gap. In principle one could state that the system becomes insulating at half filling. However, due to the large broadening of the bands, one still finds for a very large potential range a finite intensity at the Fermi-energy. It is in this range that one finds differences between the coherent potential approximation and their cluster extensions \cite{Batt06, Rowlands09}. For $V=0.55W$ one finds a new Fermi-surface topology. Around $\Gamma$ one finds an electron pocket which is related to the host band. Around $M$ one finds a hole pocket derived from the impurity band. It is note-worthy that the coherent potential approximation is not able to reproduce this effect. \cite{Batt06, Rowlands09} The non-local coherent potential approximation, for large cluster sizes should naturally exactly reproduce the current calculations, under the condition that the cluster size taken into account is large enough. For a cluster size of $2\times2$ one can see that some features of the gaping are indeed correctly reproduced. The exact potential at which the Fermi-surface topology change takes place, as well as some details of the Fermi-surface are however different. For accurate calculations it is important to take relatively large cluster sizes, within the non-local coherent potential approximation. The effect of the bath in the non-local coherent potential approximation would be relatively minor for such large clusters sizes. It is therefore relatively straight forward to extend the non-local coherent potential approximation to the current cluster calculations. For large cluster sizes both methods are equivalent.

\subsection{Self energy}

 \begin{figure*}
    \includegraphics[width=0.95\textwidth]{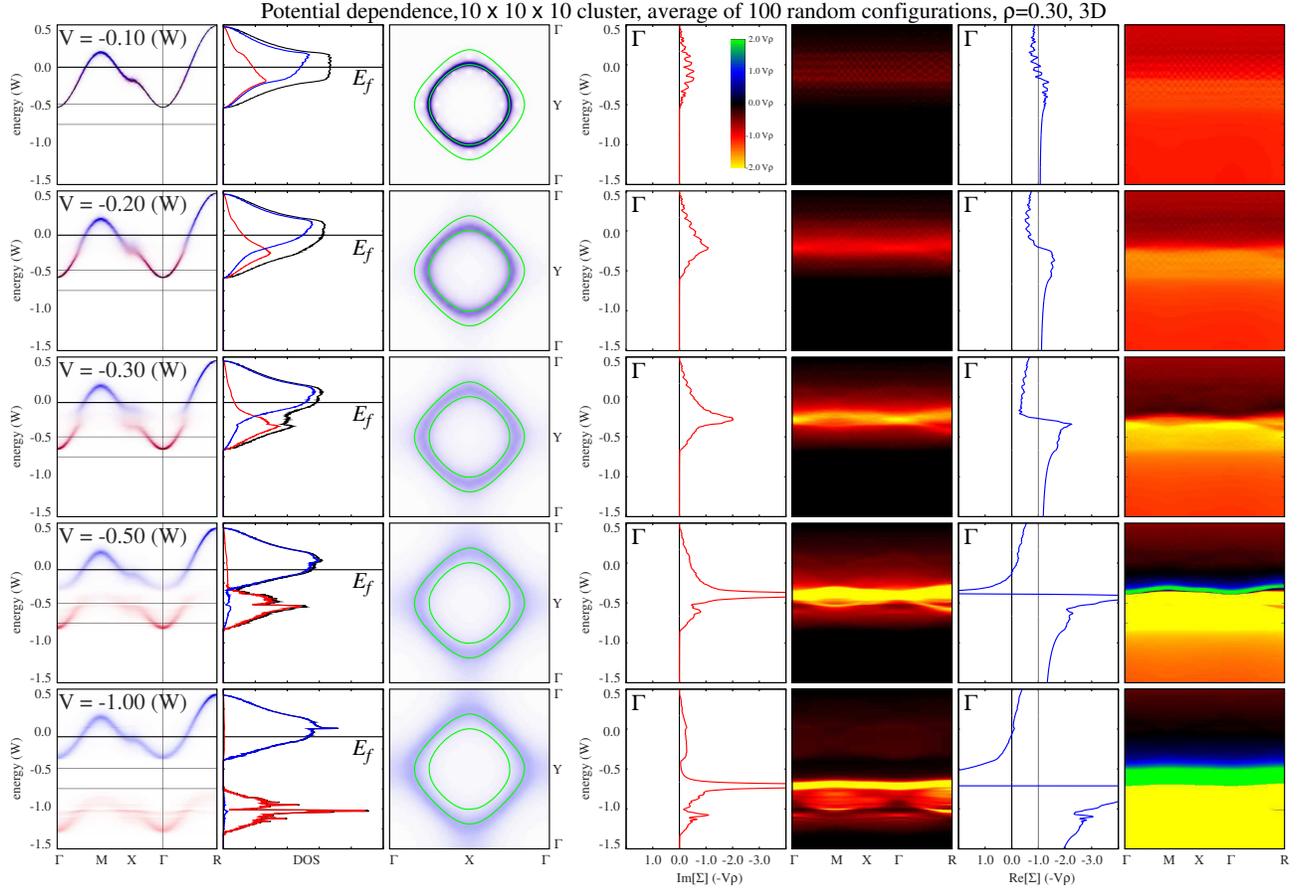}
    \caption{(color online) From left to right; the unfolded band-structure, the density of states, Fermi-surface, the imaginary part of the self energy $\Sigma$ at $\mathbf{k}=0$ ($\Gamma$), the momentum dependence of the imaginary part of the self energy, the real part of the self energy $\Sigma$ at $\mathbf{k}=0$ ($\Gamma$), and the momentum dependence of the real part of the self energy. The calculations are done for the model as defined in Eq. (\ref{eqModel}), for a three dimensional $10\times10\times10$ cluster, with $\rho=0.30$ and the potential varies from $V=-0.10$ to $V=-1.00$ for the different rows shown. The Fermi-surface plots assume a filling of on average 1 electron per atom. The additional green small lines in the Fermi-surface plot show the fermi-surface for the unperturbed system with a filling of 1 electron per atom (small around $M$) and a filling of 0.7 electron per atom (larger around $M$).}
    \label{FigPotentialDependence3D}
 \end{figure*}

A powerful concept for the understanding of random impurity calculations is the self energy $\Sigma(\mathbf{k},\omega)=\frac{1}{G_{\mathbf{k}}^{(0)}(\omega)}-\frac{1}{G_{\mathbf{k}}^{\phantom{(}}(\omega)}$. The self energy is a measure of how-much the band-structure of the system with impurities deviates from the system without substituted atoms. The full Green function is given in terms of the original Green function ($G_{\mathbf{k}}^{(0)}(\omega)$) and the self energy as:
\begin{equation}
G_{\mathbf{k}}(\omega)=\frac{1}{\frac{1}{G_{\mathbf{k}}^{(0)}(\omega)}-\Sigma(\mathbf{k},\omega)}
\end{equation}
The real part of $\Sigma$ defines an energy shift of $G_{\mathbf{k}}(\omega)$, with respect to $G_{\mathbf{k}}^{(0)}(\omega)$. The imaginary part of $\Sigma$ is related to an energy broadening of the Green-function. The Self-energy is casual, meaning that in our representation the imaginary part is strictly negative, and goes to zero as $1/\omega^2$ for large absolute values of $\omega$. The real-part is related to the imaginary part by Kramers-Kronig. Note that for a system of impurities there are no specific relations for the self-energy at the Fermi energy. As seen in Fig. (\ref{FigClusterSize}) and (\ref{FigPotentialDependence2D}) there is a substantial broadening of the Fermi surface (non-zero imaginary part of the self energy) as well as a possible change of Fermi surface topology (energy and momentum dependent real and imaginary part of the self energy). 

It is generally excepted that a system with random impurities has a broadened band-structure compared to a pure system. The imaginary part of the self energy must be integrable and goes to zero for large and small energies. By virtue of the Kramers-Kronig relations this means that one then also must have a downward energy-shift for energies lower then the main broadening and an upward energy shift for energies higher then the energy at which the broadening takes place. The broadening of bands will thus go hand in hand with shifts in the band-structure. This will become more clear with the use of the example given in Fig. (\ref{FigPotentialDependence3D}).

In Fig. (\ref{FigPotentialDependence3D}) we show the band-structure, density of states, Fermi-surface (for a filling of one electron per atom) and real and imaginary part of the self energy at $\Gamma$ as well as their momentum dependence for a $3D$ cluster as a function of the impurity potential at $\rho=0.3$. The real part of the self energy at $\Gamma$ (sixth column) has a constant part, independent of the energy, equal to $V\rho$, which one would expect in linear order in $V$. In the forth column one can see the imaginary part of the self, energy, which peaks around $V$. This leads by Kramers-Kronig to a real-part of the self-energy, which is related to a shift of the band-structure.

The shift of the band-structure due to the real-part of the self energy can be seen very clearly in the Fermi-surface plots. For a small potential of the order of $V=-0.10W$ or smaller the band-structure, density of states and Fermi-surface are, aside from some broadening, not changed much compared to the original host electronic structure. The Fermi-surface of the host is shown in the third column, by the small green pockets around $M$. For $V=-0.10W$ the Fermi-surface of the system with impurities overlaps with the original Fermi-surface. For a larger impurity potential the band-broadening becomes more substantial, and a corresponding shift of the band-structure can be seen. For the Fermi-surface this results in apparently larger pockets around $M$. The electron count does not change between the different calculations shown in the different rows in Fig. (\ref{FigPotentialDependence3D}), The Fermi-energy is defined such that there are 1000 electrons in the  cluster containing $10\times10\times10$ atoms. It appears that Luttinger's theorem is violated. Calculating the number of electrons in the system by looking at the Fermi-surface as defined as the maximum intensity of $A(\mathbf{k},\omega)$ at the Fermi energy gives an incorrect number of electrons. One can understand these findings by realizing that the calculations are not done for a periodic system, but averaged over many different randomly ordered periodic systems, within a super cell. The Green function is obtained by unfolded these super cell calculations back to the original Brillouin zone. The conditions for Luttinger's theorem to be applicable are not fulfilled.

The change in Fermi-surface can be understood in a different way. For large negative impurity potentials the substituted atoms will build an impurity band, at a lower energy then the original host band. This impurity band, will be doubly occupied, i.e. full. This doubly occupation of the impurity band removes $\rho$ electrons from the Host band, leading to an apparent change in electron occupation. In order to test this simple picture we added the Fermi-surface one would expect for an unperturbed system with $\rho=0.30$ holes, i.e. on average only 0.7 electrons per unit cell. These are the large Green Fermi-surface contours shown in the third column of Fig. (\ref{FigPotentialDependence3D}). Indeed it seems that for large potentials the apparent hole doping can be understood in terms of doubly occupied bound states around the substituted atoms, which do not contribute to the Luthinger's count. There is no clear border between the two regimes where for small potentials the virtual crystal approximation is reasonably fulfilled to a regime for large potentials where the system can be understood in terms of an impurity band. There is always some deviation from the virtual crystal approximation, however only to quadratic order in the impurity potential.

\section{\textit{ab initio} results}

There is an important difference between model calculations and self-consistent first principle studies. Within a model the potential of the host and impurity are fixed, irrespectively of their electron count. Within a self-consistent calculation the potential depends on the local electron count. This later effect will lead to screening, and different impurity potentials when two impurities are nearest neighbors compared to impurities infinitely far separated from each other. The model presented before will go astray if one places a large patch of impurity atoms next to a large patch of host atoms. In this case there will be a charge donation of the host into the impurity, independent of the distance between them, thus even if the impurity atom is very far away from the host. In reality this will cost Coulomb energy, which scales with the distance and should prevent too much of a charge flow. It is thus important to do self consistent calculations. Density Functional Theory captures these effects reasonably well. The method, language and general features described in the previous sections are directly transferable to density-functional super cell calculations.

\subsection{Cluster shape dependence for Li$_{0.75}$Mg$_{0.25}$}

 \begin{figure}
    \includegraphics[width=0.5\textwidth]{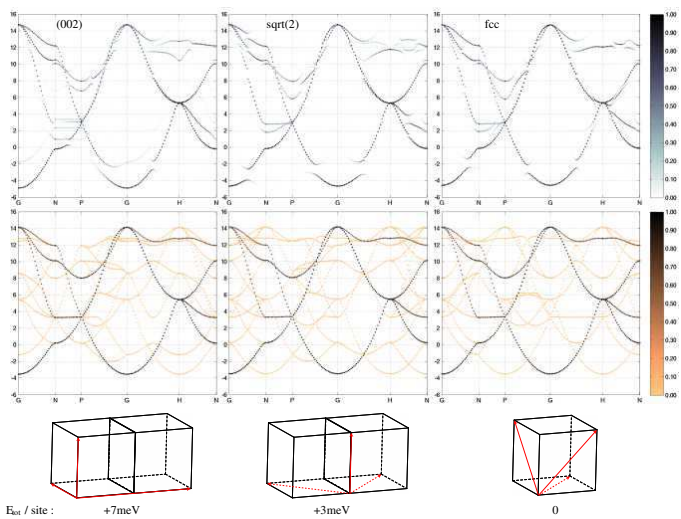}
    \caption{(color online) Calculations for: Left; a super cell of ($[100]$,$[010]$,$[002]$). Middle; a super cell of ($[110]$,$[\bar{1}10]$,$[001]$). Right; a super cell of ($[110]$,$[101]$,$[011]$). Bottom row shows the super cell and the total self-consistent LDA energy with respect to the most stable arrangement. Middle row shows the folded and un-folded LDA band-structure of pure Li within the super cell. The bands with weight zero (should disappear in the unfolding scheme) are plotted in orange, the unfolded bands (with weight 1) in Black. Top row shows the unfolded bands for the Li$_{0.75}$Mg$_{0.25}$ ordered alloy within the 3 different super cell configurations.}
    \label{FigDFTclusterform}
 \end{figure}

Mean-field approximated model calculations including several thousands of sites are possible and finite size effects therefore can be largely eliminated. Density functional theory calculations are more involved and although modern computers are very powerful it is important to try to optimize cluster sizes and shapes. Naturally Mean-field theories like most implementations of Density functional theory are still numerically easier to solve then true many-body calculations, from which some knowledge about optimal cluster sizes can be obtained.\cite{Betts99} In Fig. (\ref{FigDFTclusterform}) we show three calculations for the LDA band-structure of Li$_{4}$ (middle panels) and Li$_{3}$Mg$_{1}$ (top panels), calculated with the use of LMTO.\cite{Andersen75} The calculations all correspond to different ordered structures of the same chemical composition. The difference between the three rows is the shape of the super cell. In the middle row one can see in orange that the folded band-structure depends on the choice of the super cell used. The unfolded bands are independent of this choice (when no impurities are introduced). In the top row one can see the unfolded bands when one Li atom is replaced by Mg. If one compares the bands of Li$_{3}$Mg$_{1}$ with the folded bands of Li$_4$ then a striking feature appears. Looking for example to the lowest $s$ derived band at -4 to 0 eV going from $\Gamma$ to $N$ in the unfolded Brillouin zone. Whenever the Li$_{4}$ folded bands cross the Li$_{4}$ main bands (see middle rows) a gaping of the bands at that place in $k$ space appears when one Li atom is replaced by Mg.

This is strongly related to the difference between the calculation of an ordered structure and the calculation of a randomly substituted material. For ordered superstructures gaps appear when the folded bands cross the original bands. The dependence of this folding on the shape of the super cell opens the possibility to not only average over different configurations, i.e. different impurity densities and different orderings of the impurities, but also to average over different cluster shapes.

The beauty of \textit{ab initio} calculations over model calculations is that one for each calculation assuming a certain impurity configuration and super cell shape, not only obtains the electronic properties, but also a total energy. This total energy can be used as a weighting factor when the different cluster sizes are averaged. On the bottom of Fig. (\ref{FigDFTclusterform}) we show the total energy of the 3 different super cell calculations per site, with respect to the energy of the lowest energy state. The material studied will be grown or annealed at a certain temperature, where one can assume the system to be in thermal equilibrium. Then one could rapidly cool the system. The distribution of the different clusters can be assumed to be governed by the Bolzman statistics one would have at the annealing temperature. Annealing the LiMg crystal at 300K would give a distribution of 0.29, 0.33 and 0.38 for the three different super cells as shown in Fig. (\ref{FigDFTclusterform}). If the sample would be in thermal equilibrium at 30 K the distribution function would be 0.05, 0.23 and 0.72. Naturally the before sketched numbers are just exemplary. A real calculation needs much larger cluster sizes and the average over more then three different configurations or super cell shapes in order to give realistic results.

\subsection{Cluster size and averaging over the number of impurities in different super cells for Li$_{0.75}$Mg$_{0.25}$}

 \begin{figure}
    \includegraphics[width=0.5\textwidth]{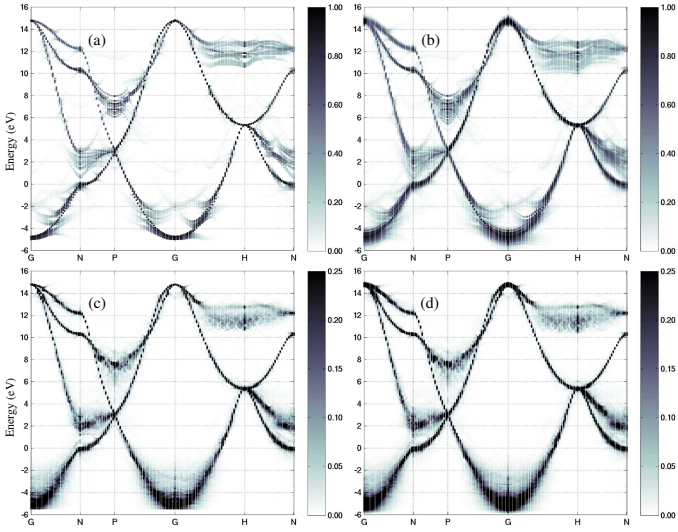}
    \caption{(color online) LDA band-structure of Li$_{75}$Mg$_{0.25}$ calculated using a super cell of size $2\times2\times2$ ($N_t=16$) for panel (a) and (b) or a super cell of size $4\times4\times4$ ($N_t=128$) for panel (c) and (d). The calculations are averaged over different configurations, whereby in panel (a) and (c) the number of Mg atoms is fixed at 4 and 32 and only the random positions varies between the different configurations, whereas in panel (b) and (d) the number of Mg atoms in the super cell ($N_s$) as well as their random positions varies. The intensities are scaled according to probabilities given by the distribution function in Eq.(\ref{EqDistributionFunction}) with $N_t=16$ ($N_t=128$) and $\rho=0.25$.}
    \label{FigDFTsizeRho}
 \end{figure}

The model calculations as shown in the previous section show averages over configurations where both the possible number of substituted atoms as well as the random placement of these atoms is varied. One would expect that for large enough cluster sizes the variation of the number of impurities becomes less important. In Fig. (\ref{FigDFTsizeRho}) we show in panel (a) a calculation done for several configurations where for each configuration the number of substituted atoms is chosen to be 4, which for a super cell of 16 sites results in a density of $\rho=0.25$. In panel (b) we show a calculation for the same super cell, but now we also vary the total number of substituted atoms. The calculation in panel (b) includes averaging over all configurations according to their distribution function as given in Eq. (\ref{EqDistributionFunction}) with $N_t=16$ and $\rho=0.25$. The average density is $\rho=0.25$ in both cases. One can see that including an average over different number of substituted atoms for small cluster sizes improves the calculation. One can compare these results to panel (c) and (d) of the same Fig. (\ref{FigDFTsizeRho}). Here the cluster size is increased from $2\times2\times2$ to $4\times4\times4$. Again in panel (c) the total number of substituted sites is kept fixed at $N_s=32$ for each configuration which varies the random positions of these substituted sites. In panel (d) also the number of substituted sites is varied between the different configurations. For larger clusters (at these impurity levels) it becomes less important to have the exact random distribution as given in Eq. (\ref{EqDistributionFunction}). One furthermore observes that for a $4\times4\times4$ cluster of a bcc structure containing $N_t=128$ sites the averaged band-structure already looks very reasonable.

\subsection{Concentration dependence in H substituted Li}

 \begin{figure}
    \includegraphics[width=0.5\textwidth]{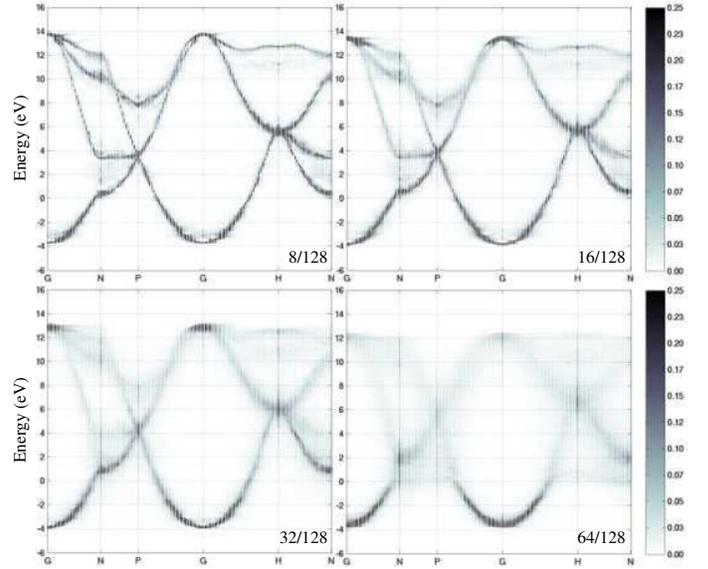}
    \caption{(color online) LDA band-structure of Li with H substitutions for different concentrations. The band-structure is calculated for a cluster of 128 atoms with random impurity configurations, unfolded to the original Brillouin zone.}
    \label{FigDFTexampleSec}
 \end{figure}

To illustrate the effect of disorder on the band structure as obtained from density functional theory, we perform DFT calculation of Li metal substituted with Hydrogen using TB-LMTO code and super cell approximation. The size of the super cell is $4\times4\times4$ (128 sites due to bcc structure). A random number generator is used to generate 100 configurations for any given H concentration. The unfolding procedure is used to present the calculated band structures in the original BZ. All 100 band structures are plotted in the same plot. Fig. (\ref{FigDFTexampleSec}) shows such plots for 6.25, 12.5, 25 and 50\% of Hydrogen. The impurity effect is twofold, first, the host band structure is broadened in both energy and momentum, and, second, there is an upward shift in energy with respect to the Fermi energy. The shifts can be understood in terms of on-site energy differences between Li and H states. Since the ionization energy of H is about 8.2eV higher than that of Li, the H $s$-states  are expected to have higher binding energy than that of Li. This means that the Hydrogen impurity is an acceptor. The shifts, however, are compensated somewhat by the changes in the band widths. At about 50 percent, the band structure of LiH alloy turns over to that of pure H which is also broadened and shows the shifts in the Fermi energy which is now due to Li$^+$ impurities as can be seen in the right bottom panel of Fig. (\ref{FigDFTexampleSec}).

\subsection{Order versus disorder in the Li$_{0.5}$ H$_{0.5}$ alloy}

 \begin{figure}
    \includegraphics[width=0.5\textwidth]{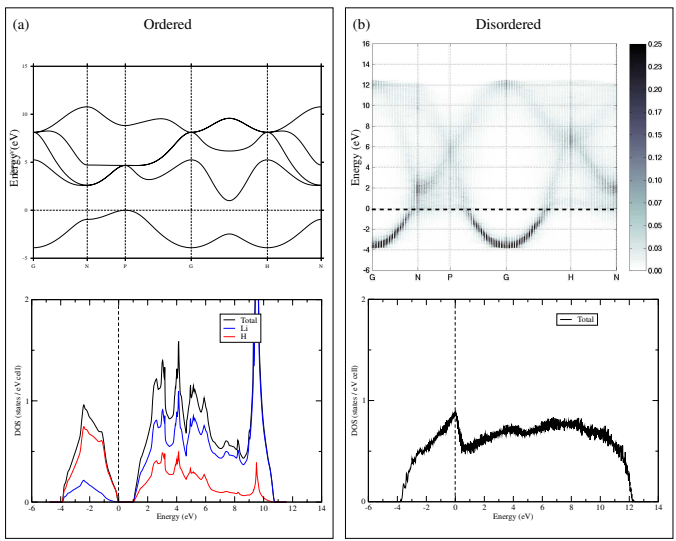}
    \caption{(color online) LDA band-structure and density of states for Li$_{0.5}$H$_{0.5}$. On the left calculations are shown for an ordered Li H structure. On the right calculations are shown for a disordered system. Calculations are done for a 128 site super cell with random impurity distributions, unfolded to the original Brillouin zone.}
    \label{FigDFTexampleThird}
 \end{figure}

The specific case of Li$_{1/2}$H$_{1/2}$ is rather interesting since it allows us to study the effect of order on the electronic structure. The electronic structure of the ordered alloy is shown in the left columns of Fig. (\ref{FigDFTexampleThird}). It is calculated with the conventional bcc unit cell that corresponds to two interpenetrating simple cubic lattices. The material is an insulator with the gap of about 1eV which separates the occupied H 1s band with the conduction band formed by the states of mainly Li character. The random alloy is, however, metallic (see right columns of Fig. (\ref{FigDFTexampleThird})). We note the very large smearing effect of the band structure especially in the region dominated by lithium $p$ based states above the Fermi energy. There is, of course, no background nor smearing of the bands in the ordered case.

\section{Conclusion}

In conclusion we have discussed a band structure based procedure which can be successfully used to describe the electronic structure of disordered alloys and randomly substituted systems. We have used a simple tight binding based model to describe the physics of the differences between minimum size super cell calculations and large super cells with random distributions of impurity atoms in the cell. In order to visualize the changes we present a simple unfolding procedure which is demonstrated to work very well indeed. We have shown that with small impurity potentials the band structure is qualitatively described by a rigid shift as in a virtual crystal approximation. However even here the bands become broadened in both energy and momentum space. This broadening becomes very large for larger impurity potentials. We also show that the Fermi surfaces are strongly smeared out for medium size impurity potentials and that the peak position of the broadened structure as a function of momentum would yield a Fermi surface determined by the crossing at an energy corresponding to the electron count i.e. the "Fermi energy" which does not agree with the Luttenger theorem size of the Fermi surface.

In order to quantify the effects of the impurity scattering effects we present the real and imaginary parts of the self energy which clearly demonstrates the various effects impurity scattering and randomness has on the electronic structure. The smearing effects at the Fermi energy can cause a semiconducting density of states in the ordered case turn into a metallic density of states. This effect is quite spectacular in the comparison of a density functional based \textit{ab initio} calculation of the electronic structure of ordered and disordered LiH. The results we show for Mg Li and Li H alloys using density functional methods demonstrate that \textit{ab initio} methods can be used which opens the door to the study of a large number of real material systems.

We would like to thank O. K. Andersen and M. Berciu for stimulating discussion. We acknowledge financial support from the Candian funding agencies NSERC,CFI, and the CRC program. This work was done within the Max Planck/ UBC funded Center for Quantum Materials at UBC and was enabled in part by the use of computing resources provided by WestGrid and Compute/Calcul Canada.

\end{document}